\documentclass[12pt]{iopart}
\usepackage{graphicx}
\usepackage{braket}
\usepackage{subfig}
\usepackage{color,soul}
\usepackage{xcolor}

\begin{document}
\title[Rubidium spectroscopy at high-pressure buffer gas conditions]{Rubidium spectroscopy at high-pressure buffer gas conditions: Detailed balance in the optical interaction of an absorber coupled to a reservoir}

\author{Stavros Christopoulos$^{1,2}$, Peter Moroshkin$^{1} \footnote{Present address: Okinawa Institute of Science and Technology, 1919-1 Tancha, Onna-son, Okinawa, 904-0495 Japan}$, Lars Weller$^{1}$, Benedikt Gerwers$^{1}$, Ralf Forge$^{1}$, Till Ockenfels$^{1}$, Frank Vewinger$^{1}$, and Martin Weitz$^{1}$}

\address{$^{1}$Institut f{\"u}r Angewandte Physik, Universit{\"a}t Bonn, Wegelerstr. 8, 53115 Bonn, Germany}
\address{$^{2}$Department of Science, American University of the Middle East, Egaila, Kuwait}

\ead{weitz@uni-bonn.de, stavros.christopoulos@aum.edu.kw}

\linespread{1}

\maketitle

\begin{abstract}
Optical spectroscopy of atoms and molecules is a field where one usually operates very far from thermal equilibrium conditions. A prominent example is spectroscopy of thin vapors, where the pump irradiation leads to a non-equilibrium distribution within the electronic structure that is well shielded from the environment. Here we describe experimental work investigating absorption and emission lines of rubidium vapor subject to a noble buffer gas environment with pressure 100 - 200 bar, a regime interpolating between usual gas phase and liquid/solid state conditions. Frequent elastic collisions in the dense buffer gas sample cause a large coupling to the environment. We give a detailed account of recent observations of the Kennard-Stepanov scaling, a Boltzmann-like thermodynamic frequency scaling between absorption and emission profiles, for both atomic and molecular rubidium species in the gaseous environment. Our observations are interpreted as due to the thermalization of alkali-noble gas submanifolds in both ground and electronically excited states respectively. Both pressure broadening and shift of the high pressure buffer gas D-lines system are determined. We also discuss some prospects, including possible advances in collisional laser cooling and optical thermometry.
\end{abstract}

\noindent {\it Keywords\/}: Laser cooling, spectroscopy of alkali atoms, high pressure buffer gas, atomic collisions, thermalization, Kennard-Stepanov relation
\section{Introduction}
\label{Intro}

System-reservoir interactions are a topic of great significance in both theoretical and experimental physics \cite{Breuer}. Despite the always present reservoir, the field of atomic and molecular physics has led to many textbook-quality scientific achievements enabled by the degree of possible decoupling of quantum systems in thin vapor or even ultra-high vacuum conditions, see e.g. advances in atomic clock science \cite{Riehle}. In the absence of a coupling to the thermal environment, an atomic emission line shape matches the absorption line shape. While in general the coupling of absorbers to the environment can lead to complex line shapes, as is the case in many solid or liquid state experiments, under certain conditions a remarkably simple thermodynamic relation between the absorption and emission is obtained, as pointed out by Kennard and Stepanov \cite{Kennard,Stepanov}, and for the case of solid state systems by McCumber \cite{McCumber}. The so-called Kennard-Stepanov relation predicts a Boltzmann-type frequency scaling $f(\omega)/\alpha(\omega) \propto \exp(-\hbar\omega/k_B T)$, where $\alpha(\omega)$ and $f(\omega)$ denote the spectral profiles of absorption and emission respectively, and $T$ is the sample temperature. The relation can be understood to originate from detailed balance in a system with a substructure in the ground and electronically excited states, respectively, with the sublevel distributions being in thermal equilibrium. The Boltzmann-type universal frequency scaling is long known to be fulfilled for some dye molecules in liquid solution \cite{Lakowicz}, and has also been observed in semiconductor systems \cite{Band}. We have demonstrated that the Kennard-Stepanov relation is fulfilled with a good accuracy for rubidium atoms in dense buffer gas environment \cite{Moroshkin} and in more recent work also for several molecular dimers, K$_{2}$, Rb$_{2}$ and Cs$_{2}$ under similar conditions \cite{Christopoulos2}. These measurements have demonstrated the validity of this thermodynamic scaling in the gaseous phase. At a typical buffer gas pressure of 100 bar, the collisional rate is of order 10$^{11}$/s, which is far above the inverse of the upper state spontaneous lifetime. As collisions between alkali atoms and noble gases are long known to be extremely elastic \cite{Speller}, the observed Kennard-Stepanov scaling in the gaseous system is well understood in terms of a thermalization of alkali-noble gas quasimolecular collisional complexes in both the electronic ground and excited states in the dense buffer gas environment \cite{Moroshkin}.

The Kennard-Stepanov relation can be used for non-contact temperature determinations. Indeed the temperature extracted from the spectroscopic data in recent molecular buffer gas experiments in good accuracy agrees with the ambient cell temperature \cite{Christopoulos2}. The Kennard-Stepanov scaling between absorption and emission is also of high relevance for recent experiments on the thermalization of photon gases in dye microcavity photon Bose-Einstein condensates \cite{Klaers2010,Marelic,greveling2018density}. Related, but in the non-equilibrium regime, the Kennard-Stepanov scaling is relevant for the operation of alkali-vapor lasers, which also build upon buffer gas broadened alkali atom filled gas cells \cite{Sell}. High-pressure buffer gas alkali cells with comparable experimental parameters as the ones used in the present work have also been used in other work of our group for the demonstration of collisional redistribution laser cooling, which  has achieved relative cooling of "macroscopic'' gas samples by optical irradiation \cite{Vogl}. The Kennard-Stepanov relation in this context is relevant for an observed asymmetry between the efficiencies of laser cooling and heating for red and blue detuning from resonance \cite{Gelbwaser}, which agrees with the intuitive expectation that to heat is easier than to cool.

In alkali gas cells, both atoms and the corresponding diatomic molecular species coexist with their ratio being determined by the cell temperature, as well understood when accounting for the molecular binding energy and phase space factors. Spectroscopic experiments in high temperature samples indeed often have both atomic and molecular spectral lines in the investigated optical wavelength range. At moderate buffer gas pressures, both atomic and molecular alkali lines have long been investigated in much detail \cite{Vdovic}.

We describe spectroscopic experiments on rubidium atoms and dimers subject to high-pressure argon buffer gas in the 100 - 200 bar range. In this regime where the linewidth of transitions due to collisional broadening is comparable to the thermal energy in frequency units, we report spectroscopic experiments incorporating both atomic and molecular lines. Moreover, we give a detailed account of recent observations of the thermodynamic Kennard-Stepanov relation in both atomic and molecular spectra, again focusing on the rubidium argon mixture gaseous system. Special emphasis is given to the experimental technique that has allowed us to investigate the absorption profile of an optically thin medium, while avoiding reabsorption effects. It relies on the measurements of the fluorescence yield and is applicable in the presence of collisional redistribution. The effect of frequent collisions between rubidium and argon atoms on the broadening and shift of the atomic $D$-lines is also investigated.

In the following, Chapter 2 gives a simple theoretical model for the Kennard-Stepanov scaling and reviews some spectroscopic data. The used experimental setup along with general spectroscopic measurements is described in Chapter 3. Subsequently, Chapter 4 describes measurements and relevant analysis on fluorescence yield, broadening and shift of atomic rubidium $D$-lines, as well as the Kennard-Stepanov scaling at high-pressure argon buffer gas conditions, and Chapter 5 presents corresponding results for rubidium dimers. Finally, Chapter 6 gives conclusions.

\section{Theoretical Model}
\label{TheoreticalModel}

\begin{figure}
\centering
\includegraphics[width=1.02\columnwidth]{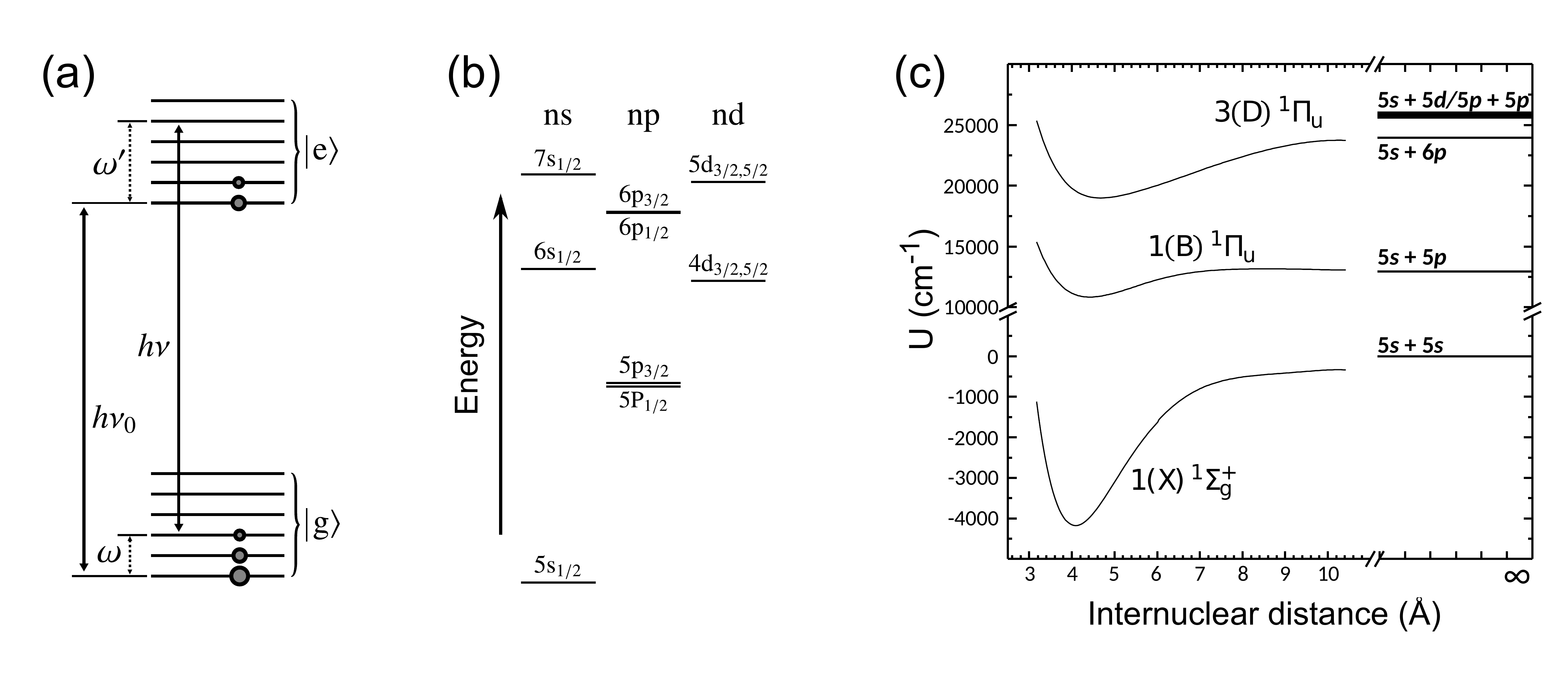}
\caption{(a) Schematic representation of an electronic ground and excited state manifold, each with an additional substructure. The size of the dots indicates the occupation of sublevels in the corresponding manifolds, each following a thermal distribution. The sublevel manifolds are coupled with an optical field of frequency $\nu$. (b) Atomic rubidium level structure. (c) Calculated molecular potentials of interest of the $\mathrm{Rb_{2}}$ dimer system \cite{Spiegelmann}, as a function of the internuclear distance. On the right hand side, the asymptotic atomic energy levels are shown. Note that both (b) and (c) refer to the level structure in the absence of a buffer gas atom.}
\label{JB}
\end{figure}

For a model of the origin of the Kennard-Stepanov relation, we here follow the treatment of Sawicki and Knox \cite{Sawicki}. Consider an electronic two-level system with a ground state $\ket{g}$ and an electronically excited state $\ket{e}$, each of which are subject to an additional sublevel structure, see Fig. 1(a). Assume that the lifetime of the electronically excited state is sufficiently long so that the sublevel distribution within the electronically excited state, as well as that within the ground state, have acquired a thermal equilibrium distribution in the presence of a coupling to the thermal environment. The occupation densities will thus be determined by the Boltzmann factors $p_g (\omega) = N_g exp(-\omega/k_B T)$ and $p_e (\omega') = N_e exp(-\omega'/k_B T)$, for the ground and excited manifold, respectively. Here, $\omega$, $\omega'$ are the energies of sublevels within the electronic ground and excited state manifold, respectively, while $N_g$, $N_e$ are normalization factors and $T$ is the temperature. For the ratio between absorption $\alpha(\nu)$ and emission $f(\nu)$ spectra we derive:

\begin{equation}
\frac{\alpha(\nu)}{f(\nu)} \propto \frac{\int g(\omega)B(\omega,\nu)p_{g}(\omega)d\omega}{\int g'(\omega')A(\omega',\nu)p_{e}(\omega')d\omega' }\propto e^{\frac{h(\nu - \nu_{0})}{k_BT}},
\label{eq1}
\end{equation}

\noindent where $g(\omega)$ and $g'(\omega')$ are the densities of state manifolds respectively. Further, $A$ and $B$ are the Einstein coefficients for spontaneous emission and absorption, respectively, and $\nu_{0}$ is the frequency of a purely electronic transition that is equivalent to a zero-phonon line in condensed matter samples (see Fig. \ref{JB}(a)). We have taken into account energy conservation in the light-matter interaction: $h \nu_0 + \omega' = h \nu + \omega$, as well as the Einstein $A-B$ relation, which here takes a form:

\begin{equation}
 g'(\omega')A(\omega',\nu)d\omega'=\frac{8\pi h \nu^{3}}{c^{2}} g(\omega)B(\omega,\nu)d\omega\,.
\end{equation}

For a small frequency interval, $(\nu-\nu_{0})\ll\nu_{0}$, the prefactor on the right hand side of the equation can be assumed constant, $h\nu^3\approx h\nu_{0}^3$ . With this, the Boltzmann-like Kennard-Stepanov scaling of Eq. \ref{eq1} is readily verified. In order to allow for a more straightforward extraction of the spectral temperature from the observed absorption and emission spectral profiles, we will later mostly use the relation in the logarithmic form:

\begin{equation}
\ln\bigg[\frac{\alpha (\nu)}{f(\nu)}\bigg] = \frac{h}{k_{B}T}\nu + D(T)\,,
\label{eq2}
\end{equation}

\noindent where $D(T)$ is a temperature-dependent (but frequency-independent) constant \cite{Neporent}. For the case of dye molecular systems, as considered in \cite{Sawicki}, the sublevel structure is given by the rovibrational levels. We are here interested in the case of alkali atoms and molecules subject to frequent noble gas collisions, which in our model can be accounted for by replacing the integration in Eq. \ref{eq1} by a summation over both bound and continuum levels of the corresponding quasi-molecular collisional manifolds. For a derivation of the Kennard-Stepanov relation in the gaseous system more closely following standard collisional theory, see \cite{Moroshkin}.

Figure 1(b) gives an overview of the lowest energy levels of the rubidium atomic system. Potential curves of the diatomic rubidium molecule as a function of the internuclear distance for the lowest electronic levels are shown in Fig. 1(c), with the asymptotic atomic energy level configuration being displayed on the right hand side. Pair interaction potential curves for the atomic Rb - Ar system that are relevant for our study of Rb atomic lines in Ar environment have been calculated \cite{Berriche}. We are not aware of any similar calculation for the $\mathrm{Rb_2}$ - Ar interaction that would be relevant for our work on $\mathrm{Rb_2}$ spectra. We point out that in the here investigated pressure regime of a few hundred bar, the impact limit gives its place to the quasi-static limit for collisions, and also multiparticle collisions become relevant.

\section{Experimental setup and broadband absorption spectroscopy}
\label{Experimentalsetup}

\begin{figure}[hb!]
\vspace*{-3em}
\centering
\includegraphics[width=1\columnwidth]{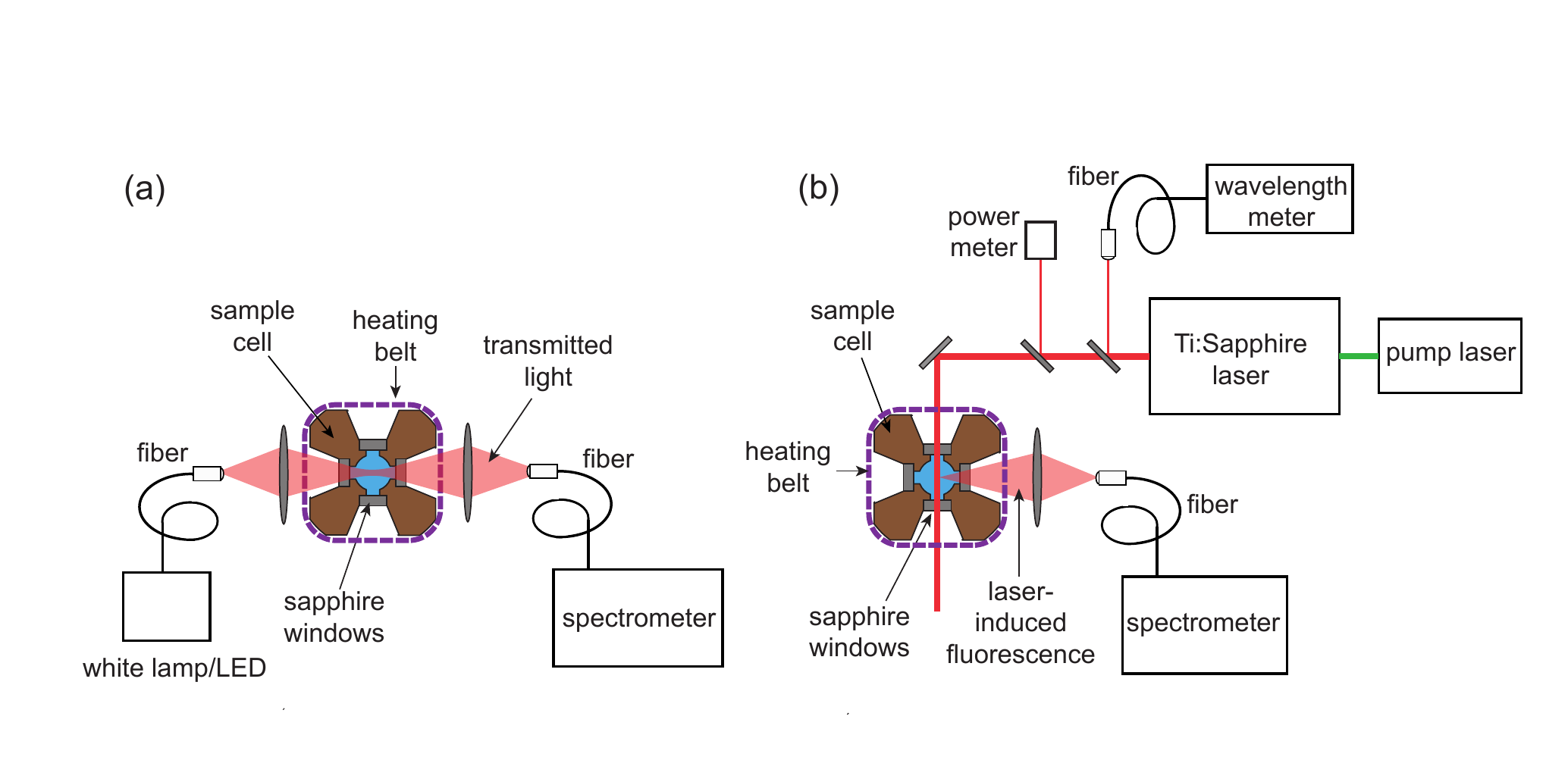}
\caption{(a) Setup for the recording of rubidium absorption spectra at high pressure argon buffer gas conditions by broadband optical irradiation. The reference spectrum is here obtained using a cell containing only argon buffer gas at the same conditions of pressure and temperature, as the respective transmission spectrum. (b) Setup for laser spectroscopic measurements of both absorption and emission lines of the high pressure buffer gas ensemble. The fluorescence was recorded in orthogonal direction with respect to the excitation laser beam to minimize the influence of the scattered radiation. Besides fluorescence spectra, also measurements of absorption spectra could be carried out with this setup by using the dependence of the  fluorescence yield  on the incident laser beam wavelength.}
\label{Setup}
\end{figure}

Figures \ref{Setup}(a) and (b) depict the spectroscopic setups utilized for the experimental investigation of thermalization of both atomic and molecular states of rubidium, which employ high pressure stainless steel cells, equipped with sapphire windows to allow for optical access. The inner volume of these cells is typically $2 - 5$ $\mathrm{cm^{3}}$, with the gas-filled spacing between the optical windows along the two orthogonal, spectroscopically used axes being approximately $1 - 1.5$ $\mathrm{cm}$. At the bottom of the cell a reservoir contains a few grams of metallic rubidium. High-pressure argon gas (up to 250 bar) is admitted through a gas inlet, while a manometer allows for pressure monitoring. The necessary optical density of the alkali vapor is reached by raising the temperature to values up to 730 K, achieved with the use of heating belts which are wrapped around both the cell and its reservoir. 

\begin{figure}[ht!]
\centering
\includegraphics[width=1\columnwidth]{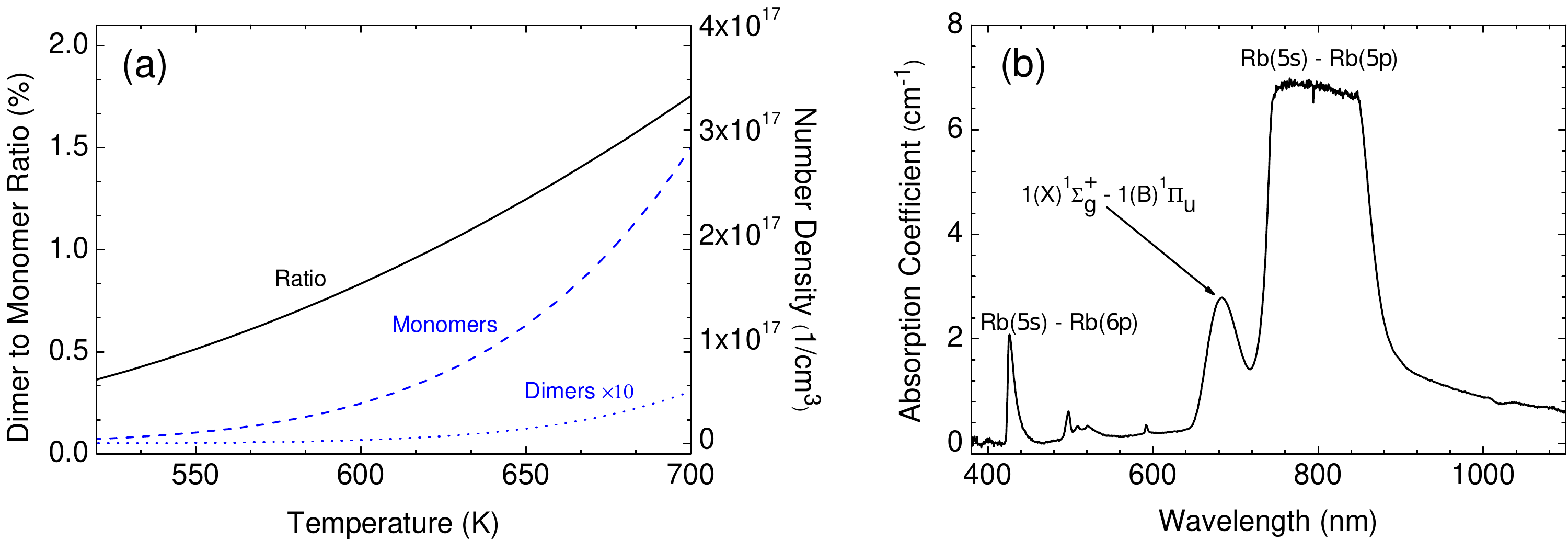}
\vspace*{-0.5em}
\caption{(a) The number densities of dimers$\time10$ (blue dotted line) and monomers (blue dashed line) of Rb and their ratio (solid black line) versus temperature in the saturated Rb vapor. (b) Experimentally determined absorption coefficient of rubidium vapor for 180 bar of argon buffer gas at a temperature of 670 K with some of the observed atomic lines and molecular bands indicated. The data was recorded by spectrally resolving the cell transmission subject to broadband irradiation of a halogen lamp.}
\label{Ratio}
\end{figure}

The alkali vapor density increases significantly with temperature, and in Fig. \ref{Ratio}(a) we give predictions for the corresponding variation of the vapor density for both Rb monomers and dimers, reaching e.g. densities of $1.8\times 10^{16}$ $\mathrm{ cm^{-3}}$ at 570 K for the atoms (see Fig. \ref{KSatom}) and $2.4\times 10^{15}$ $\mathrm{ cm^{-3}}$  at 673 K for the dimers (see Fig. \ref{KSdim}), respectively. Their density ratio favors dimers over monomers for increasing temperature, while thermal destruction of dimers is expected to begin at temperatures over 700 K \cite{Vdovic}. It is also noted that the number density of argon atoms is of the order of $10^{21}$ $\mathrm{ cm^{-3}}$ for buffer gas pressures of 200 bar, approaching within an order of magnitude the density of typical fluids.

To simultaneously verify the presence of atomic and molecular rubidium transitions, we carried out a broadband spectral absorption measurement on the high-pressure buffer gas ensemble using a setup as shown in Fig. \ref{Setup}a. The sample was irradiated by a broadband halogen lamp producing optical emission of sufficient spectral width to excite both atomic and molecular excitation lines in the visible and near infrared. Figure \ref{Ratio}(b) gives corresponding experimental data for a 670 K cell temperature for a vapor-density limited rubidium density and 180 bar of argon buffer gas pressure. The shown absorption coefficient data was obtained by collecting both reference and transmission spectra in forward direction for normalization. Due to the - within the range of our experimental sensitivity - complete absorption of light in the wavelength range of $750 - 850$ nm for the used cell temperature the actual line shape of the atomic $D_{1}$, $D_{2}$ doublet could not be resolved. The maximum observed value of the absorption coefficient of near $7$ per cm is here purely instrumentally limited. The weaker atomic Rb($6p$) $\rightarrow$ Rb($5s$) transition of the second principle series can be seen at 420 - 430 nm. In the range 650 - 730 nm an absorption peak due to the $1(\mathrm{X}){}^1\Sigma_{\mathrm{g}}^{+} - 1(\mathrm{B}){}^1\Pi_{\mathrm{u}}$ molecular Rb$_{2}$ transition is observed, whose observation verifies the presence of the molecular dimers in the dense buffer gas environment, see also earlier work in the dilute gas regime \cite{Vdovic}. In the range of 800 - 1100 nm the presence of the molecular $1(\mathrm{X}){}^1\Sigma_{\mathrm{g}}^{+} - 1(\mathrm{A}){}^1\Sigma_{\mathrm{u}}^{+}$ band is not clearly resolved in the data of Fig. \ref{Ratio}(b), as attributed to its overlap with residual wings of the strong atomic $D$-lines.

To record both absorption and emission spectra of a specific atomic/molecular resonance a tunable cw Ti:sapphire laser is employed (see Fig. \ref{Setup}(b) for the used experimental setup). Its emission is collimated to a 3 mm beam diameter, while the power is limited to 1 W in order to suppress redistributional laser cooling or heating effects. In order to reduce light scattering from the excitation beam, the fluorescence is collected in a perpendicular direction and subsequently guided via a multimode fiber to a spectrometer. The methods to measure the atomic and molecular absorption bands vary according to experimental conditions and are explained in detail below.

\section{Laser spectroscopic absorption and emission measurements of collisionally broadened atomic rubidium resonances}
\label{Thermo1}

Our investigation in atomic samples is focused on the regime where the $D_1$ and $D_2$ rubidium resonances are strongly broadened due to the frequent collisions with the buffer gas atoms. At a high atomic rubidium density (high temperature), there is a significant reabsorption of the laser-induced atomic fluorescence by ground-state rubidium atoms which mostly affects the two peaks and leads to their diminishing as compared to the signal in the wings.
Hence, for the spectroscopy measurement described in the following, we operated the cell (with an optical path of $l$) at a low optical density ($OD =-ln\big({\frac{I_{out}}{I_{in}}\big)}=\alpha \times l < 0.35$) by correspondingly regulating the cell temperature. 

\begin{figure}[hb!]
\vspace*{-11.5em}
\centering
\includegraphics[width=0.95\columnwidth]{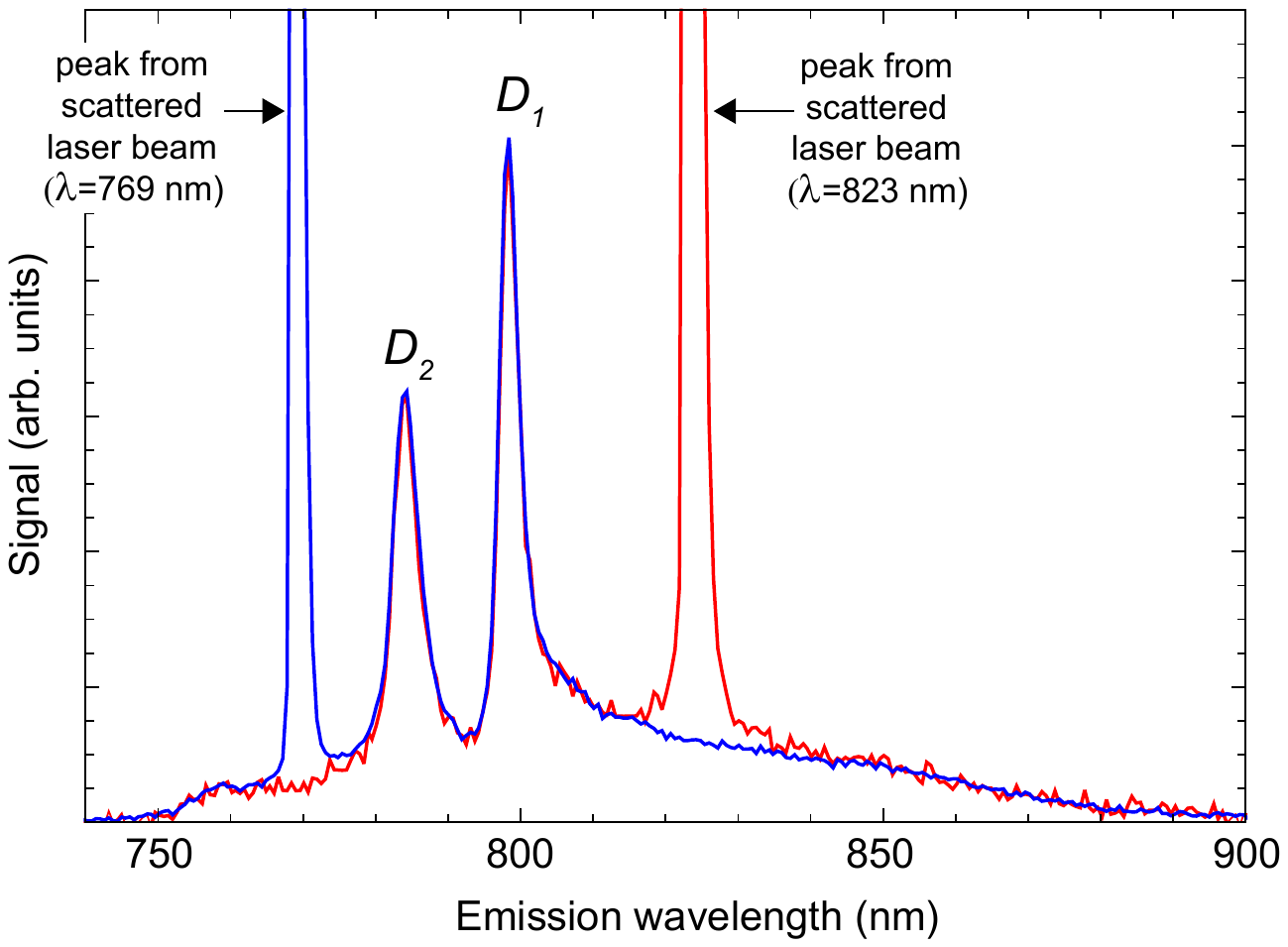}
\vspace*{-20.5em}
\caption{Spectral matching of emission profiles under excitation at 769 nm (blue line) and 823 nm (red line) for 160 bar of argon buffer gas at a temperature of 570 K. The atomic vapor density here is expected to be $1.8\times 10^{16}$ $\mathrm{ cm^{-3}}$, while the argon buffer gas density is approximately $10^{21}$ $\mathrm{ cm^{-3}}$. It is clearly observed that the emission to good accuracy is independent from the excitation wavelegth. The visible strong peaks at the exciting laser beam frequency, respectively, is due to scattered light.}
\label{Fluor2}
\end{figure}

Figure \ref{Fluor2} shows typical raw data of the emission spectrum for two different  wavelengths of the exciting laser beam. Besides a pressure broadened rubidium $D$-lines  spectrum, we observe in each spectrum a strong peak due to scattered light originating from the excitation laser beam. To exclude this effect, composite fluorescence profiles were constructed by superimposing two experimentally recorded fluorescence spectra with different excitation wavelengths. Such a procedure was possible given the observed spectral redistribution. One observes in Fig. \ref{Fluor2} that the form of the spectrum  is, to good accuracy, independent of the excitation wavelength. This is attributed to the frequent collisions of the buffer gas ensemble, causing a thermalization within the quasimolecular collisional manifolds in the electronically excited state. Indeed, this observation is an example of Kasha's rule, an effect well known in fluorescent spectroscopy of dye molecules \cite{Lakowicz}. The two obtained spectra are rescaled in intensity such that the two overlap, and the composite profiles are generated by combining the parts that do not include the scattered laser beam peak. Corresponding matched profiles are shown by the red solid line of Fig. \ref{KSatom}(a) for the case of 160\,bar of argon buffer gas pressure and 570 K cell temperature.

\begin{figure}
\vspace*{-2em}
\centering
\includegraphics[width=0.87\columnwidth]{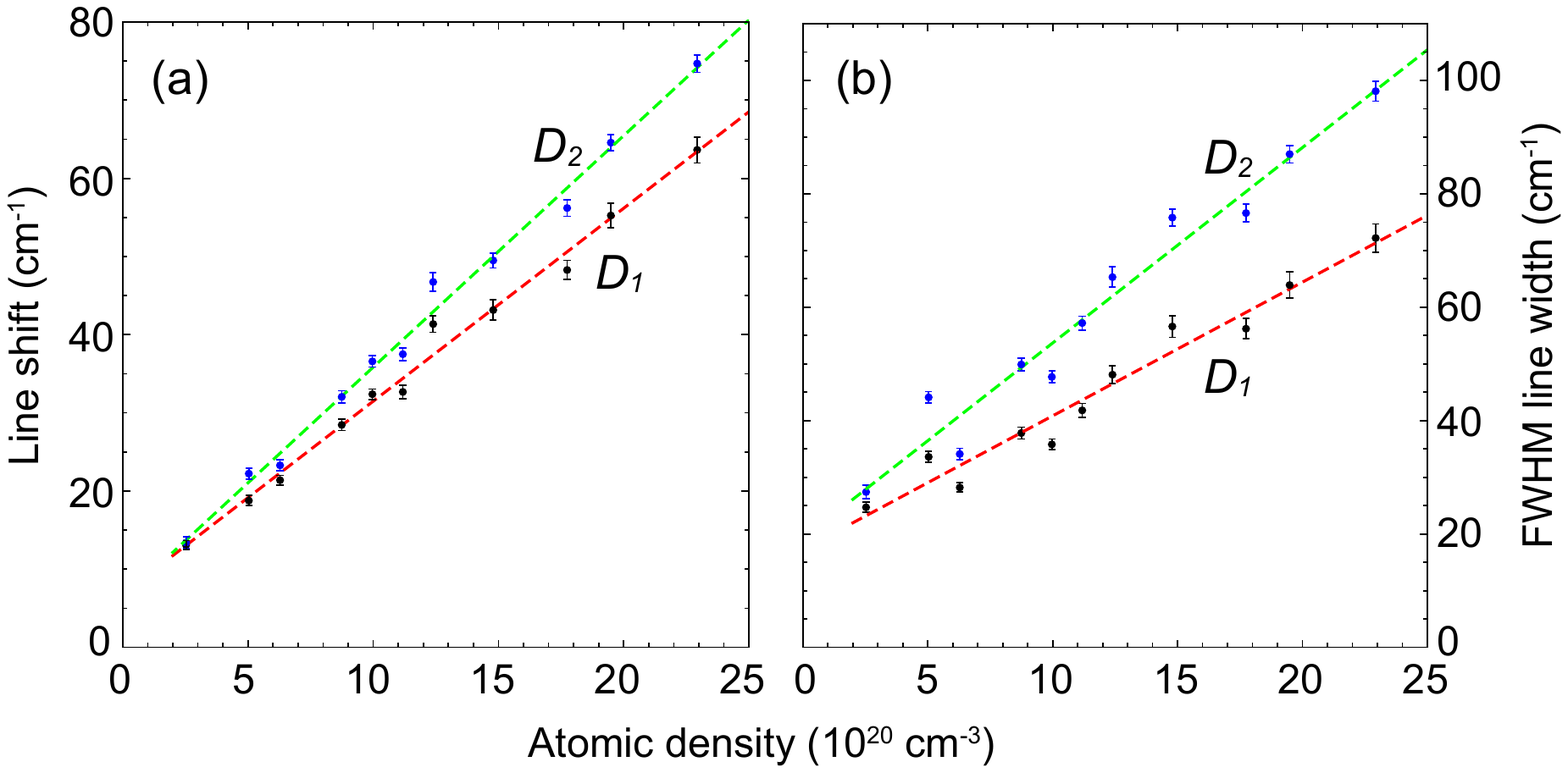}
\vspace*{-27.5em}
\caption{(a) Line shift of the $D_{1,2}$ rubidium resonances extracted for different argon buffer gas densities. The data points, along with the error bars, were obtained by fitting Lorentzian curves to the experimental fluorescence spectra. The red and green lines are linear fits on the $D_{1}$ and $D_{2}$ data, respectively. (b) Corresponding FWHM line widths of the spectra, as obtained with the same technique. The measurements were performed at a temperature of 570 K, and the density variation was obtained  by using Ar pressures in the range of $20 - 190$ bar.}
\label{FlvsDen}
\end{figure}

We next investigate the influence of frequent collisions between rubidium and argon buffer gas atoms on the emission spectra of atomic rubidium resonances. Previous works \cite{rotondaro1997collisional,ottinger1975broadening} have mostly investigated pressure broadened fluorescence spectra at much lower buffer gas pressures than was used in our work. We are also aware of pioneering high pressure absorption spectroscopy work by Ch'en et al., although being carried out at slightly varying temperature conditions \cite{Chen1,Chen2}. To analyze the dependence of linewidth and pressure shift in our experiment, we have recorded fluorescence spectra in the vicinity of the $D_{1,2}$ atomic rubidium resonances for a wide range of argon buffer gas pressures which are subsequently analyzed using Lorentzian  fits to extract both the line shifts and widths of the $D_{1,2}$ peaks, with the corresponding fit errors. In our analysis we use only the central part of each resonance, thus ignoring the wings, since the latter produce a large non-Lorentzian spectral pedestal. The error is dominated by systematic effects (which are also included in the shown error bars) due to asymmetry in the lineshape, i.e the resonance peaks exhibit significantly steeper blue than red wings. This is expected to mildly affect the determination of the line centers, resulting in slightly red shifted fit values, as well as lead to a systematic uncertainty in the deduction of the linewidth. Our obtained experimental data for the pressure shift, see Fig. \ref{FlvsDen}(a), shows a shift of both the $D_{1}$ and $D_{2}$ lines towards longer wavelength with increased buffer gas density. The extracted shift is $2.5 \pm 0.2$ $\mathrm{cm^{-1}}/10^{20}$ $\mathrm{cm^{-3}}$ for the $D_1$ resonance and $3.0 \pm 0.2$ $\mathrm{cm^{-1}}/10^{20}$ $\mathrm{cm^{-3}}$ for the $D_2$ resonance, respectively, which is in good agreement with previous work by Ch'en \cite{Chen2}, who stated that the shift is very similar for both components and approximately equals $2.6$\,cm$^{-1}/10^{20}$\,cm$^{-3}$. Our measurements show a clearly larger pressure shift for the $D_{2}$ than for the $D_{1}$ line, which is also visible in \cite{Chen2} when inspecting the data plotted. We attribute the visible non-vanishing offset of the zero-density extrapolation curve to be due to the used procedure of fitting the (central parts of the) asymmetric lineshapes with Lorentzian curves.

Figure \ref{FlvsDen}(b) shows dependence of the linewidth of the $D$-lines resonances on the argon density, also obtained with the described analysis technique. As expected, the linewidth of the resonances increases with  buffer gas density. As described above, we expected systematic uncertainties in this case to be significantly larger than for the lineshifts. We note that the non-vanishing zero-density extrapolation value here however is close to the instrumental resolution of the used spectrometer. From our fit, we derive a linewidth change of  $2.3 \pm 0.4$ $\mathrm{cm^{-1}}/10^{20}$ $\mathrm{cm^{-3}}$  for the $D_1$ resonance and $3.4 \pm 0.4$ $\mathrm{cm^{-1}}/10^{20}$ $\mathrm{cm^{-3}}$ for the $D_2$ component, in good agreement with the work by Ch'en ($2.3$ $\mathrm{cm^{-1}}/10^{20}$ $\mathrm{cm^{-3}}$ and $3.2$ $\mathrm{cm^{-1}}/10^{20}$ $\mathrm{cm^{-3}}$ for the $D_1$ and $D_2$ component, respectively). The error bars of our quoted values are dominated, besides systematic uncertainties, by the resolution of the used spectrometer.

\begin{figure}
\centering
\includegraphics[width=0.6\columnwidth]{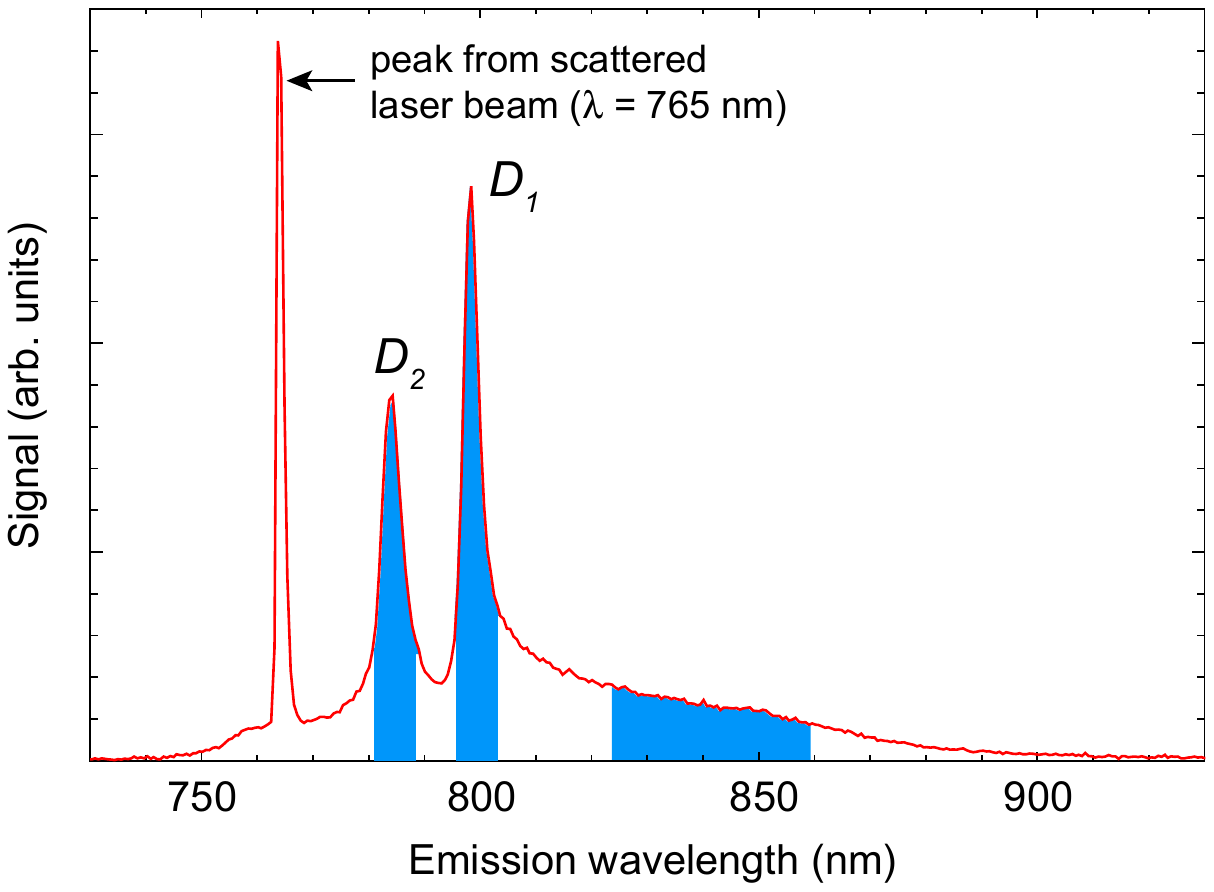}
\caption{Rubidium fluorescence spectrum (red line) in the vicinity of the atomic $D$-lines for excitation wavelength of 765 nm, at 570 K cell temperature and 160 bar argon buffer gas pressure. 
The integrated area between 824 nm - 860 nm is used to produce the respective fluorescence yield. The areas under the  $D_1$ and $D_2$ resonances are used for cross-reference. 
}
\label{Yield}
\end{figure}

While in the emission spectra the wings are clearly visible, the absorption profile, for the here discussed low optical density case, exhibits wings that are too small for accurate measurements directly from the spectrally resolved transmission of a broadband source as done for the data in Fig. \ref{Ratio}(b). The absorption data shown in Fig. \ref{KSatom}(a) (blue dashed line) was thus obtained by conducting measurements of the fluorescent yield obtained by irradiation of the rubidium cell with  the tunable Ti:sapphire laser. For such a low optical density sample, perturbing reabsorption effects are expected to be small. It is known that collisions between electronically excited alkali atoms and noble gas atoms are extremely elastic \cite{Speller,Vogl}, so in the alkali-noble buffer gas system quenching effects are significantly suppressed. The absorption coefficient is thus expected to be proportional to the fluorescence yield. The absorption spectral profile could correspondingly be determined by conducting excitation wavelength dependent measurements of the fluorescence yield. Experimentally we conducted excitation-dependent fluorescence measurements for a broad range of excitation wavelength using the setup shown in Fig. \ref{Setup}(b). In principle, the fluorescence yield can be obtained for each of the investigated wavelength values by integrating over the entire obtained corresponding spectrum. Experimentally however, scattered light at the incident laser beam wavelength, see Fig. \ref{Yield} for a typical obtained spectrum, would lead to miscalculations. At this point, we take advantage of the spectral redistribution of fluorescence due to the frequent collisions in the dense buffer gas system, which makes the emission spectrum in good accuracy independent of the excitation wavelength (see also Fig. \ref{Fluor2}). Thus the fluorescent yield is determined by integration over a reduced region of the fluorescence spectrum, chosen to be offset from the corresponding excitation laser wavelength, as indicated in Fig. \ref{Yield}. To validate our approach the area underneath the $D_1$ and $D_2$ resonances is checked for reference. Finally, plotting the integrated emission power versus the excitation wavelength allows the reconstruction of the absorption spectrum.

\begin{figure}
\vspace*{-3.5em}
\centering
\includegraphics[width=1\columnwidth]{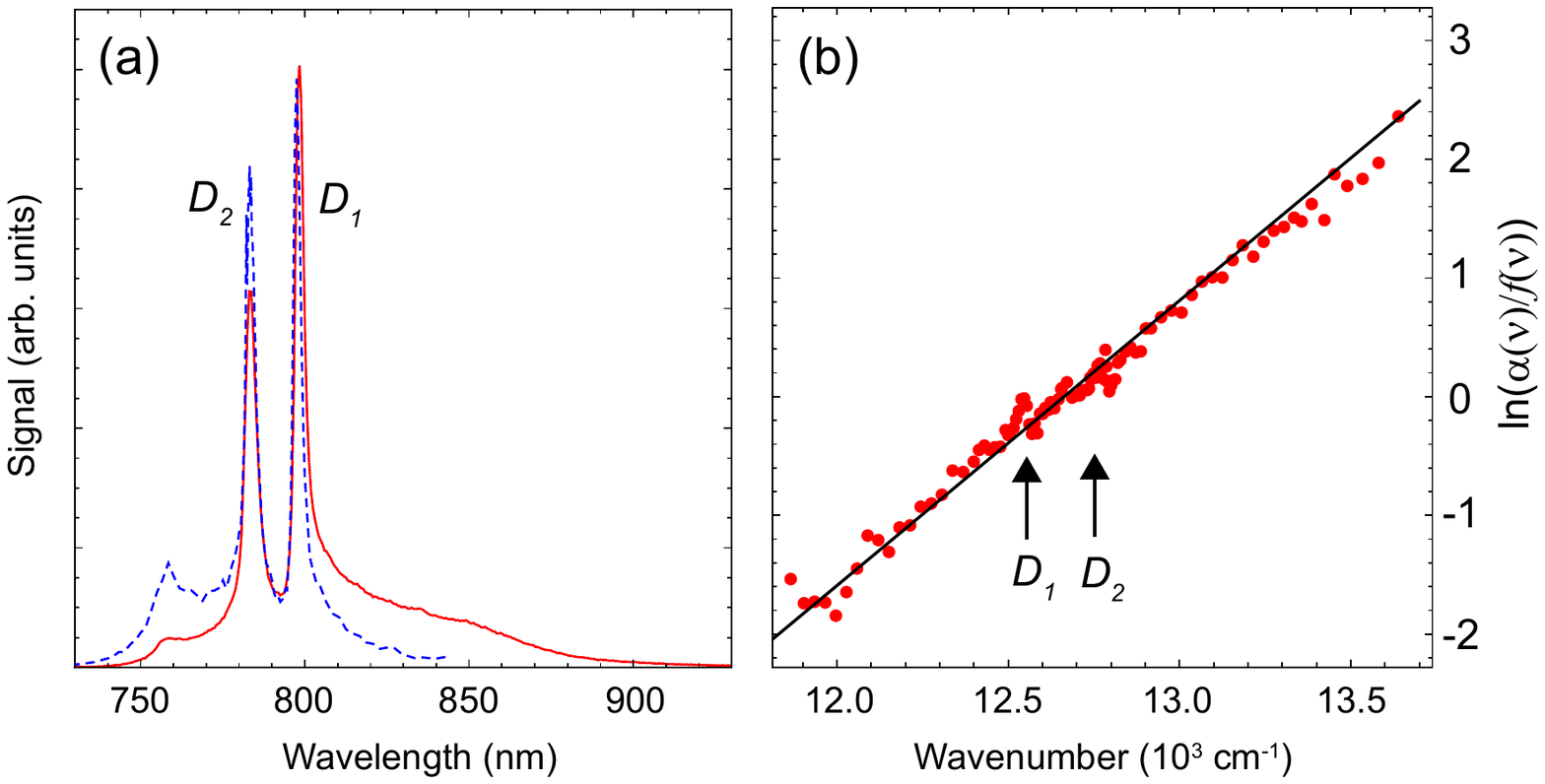}
\vspace*{-33.5em}
\caption{(a) Rubidium emission spectra (red solid line) and absorption (blue dashed line) in the vicinity of the atomic rubidium $D$-lines at 570 K cell temperature for 160 bar argon buffer gas pressure. The atomic vapor density here is expected to be $1.8\times 10^{16}$ $\mathrm{ cm^{-3}}$, while the argon buffer gas density is approximately $10^{21}$ $\mathrm{ cm^{-3}}$. (b) The red dots give the logarithmic ratio of the absorption and emission spectral profiles as a function of the optical wavenumber. The black solid line has a slope $hc/k_B T$, with $T=579$ K corresponding to the temperature of the heated gas cell.}
\vspace*{1.5em}
\label{KSatom}
\end{figure}

Figure \ref{KSatom}(a) shows the absorption (blue dashed line) and emission (red solid line) spectra acquired at a temperature of 570 K, for argon buffer gas pressure of 160 bar, see our earlier work \cite{Moroshkin} for data at other buffer gas pressure and temperature values. The spectra show relatively sharp peaks, corresponding to the $D_{1}$ and $D_{2}$ lines, on top of a spectrally broad pedestal. Furthermore, an additional spectroscopic feature, visible mainly in the absorption spectrum and centered at approximately 750 nm, can be attributed to the complex form of the pseudo-molecular potentials at small distances between rubidium and argon pairs. We have fitted the central parts of both resonances with Lorentzian lineshapes also for the absorption spectral data. The resulting lineshifts and linewidths are in agreement with both the data from \cite{Chen1,Chen2} and with our emission spectra measurements discussed above. 

Figure \ref{KSatom}(b) shows the logarithmic ratio (red dots) between absorption and emission spectra versus the optical wavenumber, where to good accuracy a linear behavior is found even in the vicinity of the strong $D$-lines, as predicted by the Kennard-Stepanov relation. We fit the linear region of the logarithmic ratio $\ln\big(\frac{\alpha (\nu)}{f(\nu)}\big)$ (black solid line), and extract the slope of the fit along with its corresponding uncertainty. In accordance to Eq. \ref{eq2}, the slope is equal to $hc/k_B T$, producing a spectral temperature of $T_{extr}=579(7)$ K, which is in a good agreement with the measured cell temperature of 570 K. Deviations from the linearity of the data are observed towards the far blue and red spectral areas, where instrumental limitations on the already low signal become apparent. Our observations demonstrate that the Kennard-Stepanov relation is well fulfilled in the high-pressure buffer gas regime, as attributed to the frequent collisions between rubidium and argon atoms leading to thermalization of the pseudo-molecular manifolds in both the ground and excited states. It is noted that measurements conducted at significantly lower buffer gas pressures show a very limited linear behavior of the Kennard-Stepanov scaling, which supports the assumption of the collisions with the buffer gas atoms being responsible for the thermalization within the quasimolecular collisional manifolds.

\section{Kennard-Stepanov spectroscopy of molecular rubidium resonances}
\label{Thermo2}

\begin{figure}[ht!]
\vspace*{-2.5em}
\centering
\includegraphics[width=0.6\columnwidth]{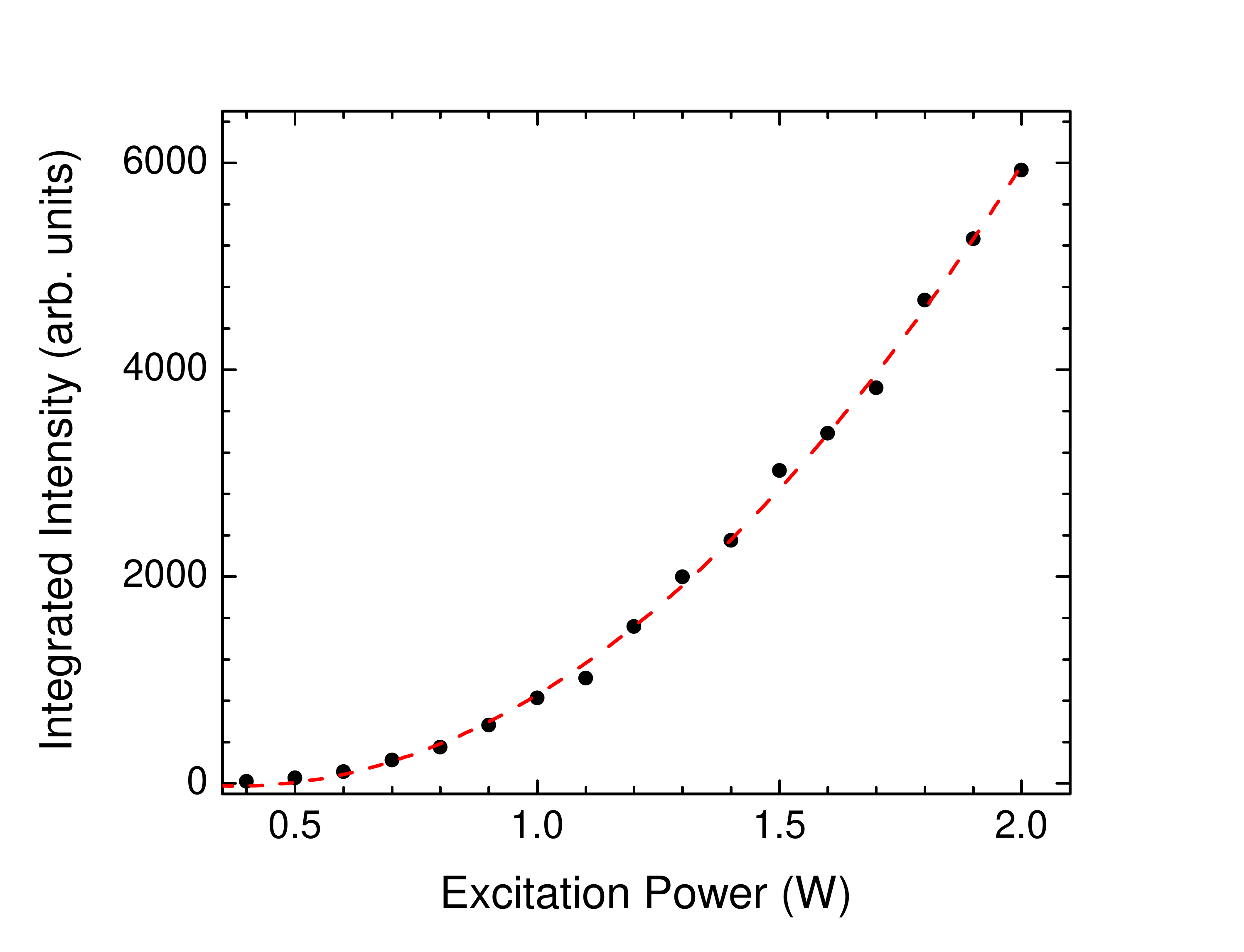}
\caption{Integrated fluorescence (black circles) of the Rb($6p$) $\rightarrow$ Rb($5s$) atomic transition following non-resonant excitation by a 780 nm wavelength laser beam, measured for different excitation powers, along with a quadratic fit (red dashed line). The data are taken at a temperature of 673 K for 150 bar argon buffer gas pressure.}
\label{PD}
\end{figure}

Our study continues with the investigation of the Kennard-Stepanov relation in molecular bands of the rubidium dimer, also subject to high-pressure noble buffer gas conditions, as described above. More specifically, we present recent results involving the $1(\mathrm{X}){}^1\Sigma_{\mathrm{g}}^{+} - 3(\mathrm{D}){}^1\Pi_{\mathrm{u}}$ electronic transition, see also Fig. \ref{JB} (c). We irradiate the heated rubidium cell with a laser beam of wavelength between 770 and 830 nm tuned in the vicinity of the rubidium $D$-lines. As discussed above, in the high-pressure buffer gas sample, excitation of the atomic Rb$(5p)$ states is possible  also by far detuned radiation via collisional redistribution. Excitation of the $3(\mathrm{D}){}^1\Pi_{\mathrm{u}}$ state is energetically possible following the absorption of two photons. It can occur through energy pooling effects involving rubidium atoms and dimers as well as argon atoms. Following the collision of two optically excited Rb($5p$) atoms, energy pooling in the high-pressure ensemble can populate the atomic Rb($6p$) and Rb($5d$) states. Power dependence measurements of the integrated fluorescence conducted under non-resonant conditions are shown in Fig. \ref{PD}, where the clear quadratic dependence  is understood from the two-step excitation dynamics. The molecular $\mathrm{Rb}_2$($3(\mathrm{D}){}^1\Pi_{\mathrm{u}}$) state can then be populated through the collision of rubidium Rb($6p$) or Rb($5d$) atoms with either ground state atoms or ground state rubidium molecules, see our earlier work \cite{Christopoulos2} for details.

\begin{figure}[hb!]
    \centering
\vspace*{-0.5em}
\includegraphics[width=0.9\columnwidth]{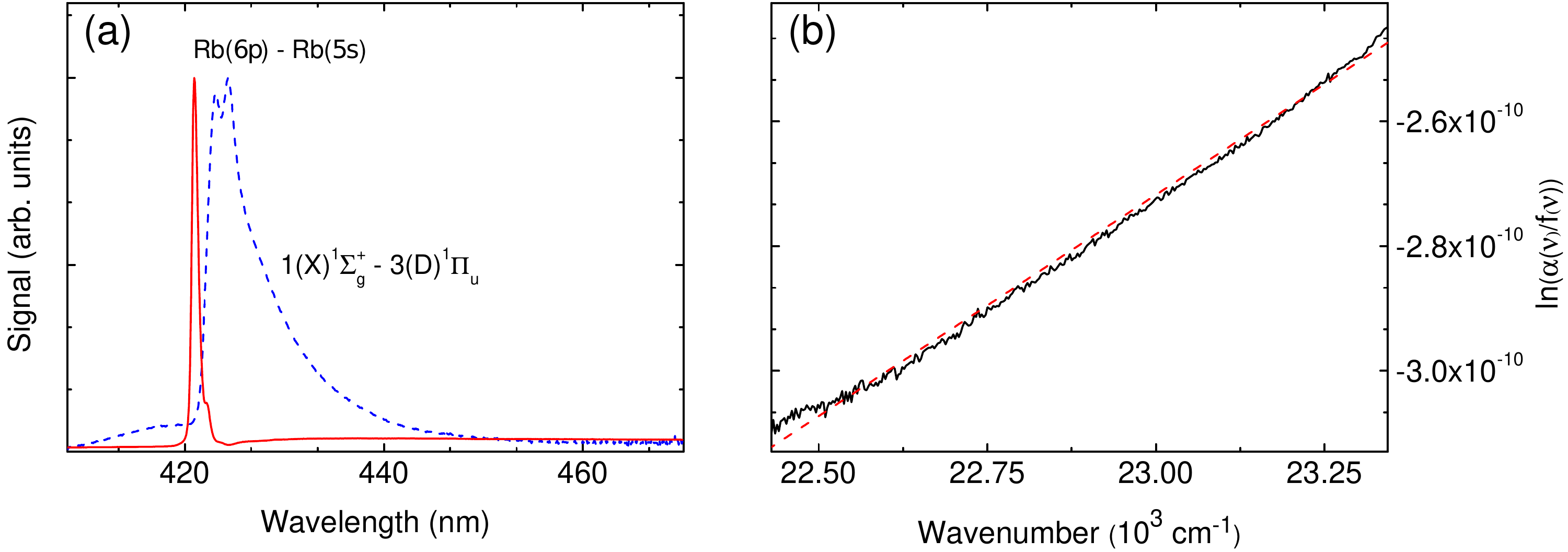}
    \caption{(a) Experimental absorption (blue dashed line) and emission (red solid line) spectra of rubidium vapor subject to 150 bar argon buffer gas at 673 K cell temperature. The dimer vapor density here is expected to be $2.4\times 10^{15}$ $\mathrm{ cm^{-3}}$, while the argon buffer gas density is approximately $10^{21}$ $\mathrm{ cm^{-3}}$. The molecular contribution at the red detuned side of the atomic resonance is visible in both emission and absorption (in the emission, the molecular fluorescence starting above approximately 430 nm wavelengh appears significantly suppressed in comparison to the visible strong atomic second principle series line). The emission spectra were recorded by irradiating the sample with a laser beam near 800 nm wavelength, and the absorption data is the spectrally resolved absorption using a blue light-emitting diode. (b) Logarithmic ratio (black solid line) of absorption and emission spectral profiles as a function of optical wavenumber. The red dashed line is a linear fit.}
\vspace*{-1em}
    \label{KSdim}
\end{figure}

Due to both the lower densities and the lower oscillator strength of the molecular transitions, the corresponding optical densities are significantly lower than those of the earlier described atomic transition measurements, thus suppressing reabsorption effects, even at higher cell temperatures. Figure \ref{KSdim}(a) shows typical experimental absorption (blue dashed line) and emission (red solid line) spectral profiles for the rubidium dimer $1(\mathrm{X}){}^1\Sigma_{\mathrm{g}}^{+} - 3(\mathrm{D}){}^1\Pi_{\mathrm{u}}$ transition acquired for 150 bar argon buffer-gas pressure and 673 K cell temperature. The absorption profile in this case was obtained by irradiating the sample with a blue light-emitting diode (LED) and collecting both reference and transmission spectra in forward direction for normalization, as shown in Fig. \ref{Setup}(a). Both spectral profiles exhibit the sharp atomic doublet Rb($6p$) $\rightarrow$ Rb($5s$) over the broad dimer band, the latter being comparatively weak in the observed emission spectrum. 

Figure \ref{KSdim}(b) shows the logarithmic ratio of absorption and emission (black solid line), together with a linear fit (red dashed line). The measurements show that the Kennard-Stepanov relation is also found to be well fulfilled for a molecular electronic transition, as attributed to the thermalization of rovibronic manifolds in the upper and lower states due to the frequent collisions in the dense buffer gas sample. The slope in this case yields a spectral temperature $T_{extr} = 677(6)$ K, in good agreement with the measured cell temperature. A study of the thermalization in other alkali dimer manifolds has been reported in previous work of our group \cite{Christopoulos2}.

\section{Conclusions and Outlook}

At a buffer gas pressure of a few hundred bars, the pressure broadened linewidth of alkali absorption lines reaches, or even exceeds, the thermal energy in frequency units, and interpolates between the usual sharp atomic physics spectral lines and the band structure of condensed matter systems. In this regime, effects of the thermalization on the spectral line shapes can be studied in gaseous ensembles. In this work, we have determined pressure broadening and pressure shift of dense alkali-atom noble gas mixtures. Atomic and molecular resonances have been observed that in good accuracy fulfill the Boltzmann-type Kennard-Stepanov scaling between absorption and emission spectral profiles, as is attributed to the thermalization of alkali-buffer gas collisional manifolds, as well as rovibronic manifolds of alkali dimer molecules. The spectra obtained at these high pressures clearly show that the environment does play a vital role also in gaseous systems. Indeed, the observation of the Kennard-Stepanov scaling shows that collisions are frequent enough to thermalize the quasimolecular collisional complexes manifolds, and thus cannot be treated as a small perturbation.

For the future, the dense gas system can serve as a model for system-reservoir interactions. Other prospects include non-contact, temperature measurements from the Boltzmann-type Kennard-Stepanov frequency scaling, or even atom- (or molecule) based determinations of the Boltzmann constant. Non-contact, spectroscopic temperature determinations are highly relevant also for studies of collisional redistribution laser cooling on the dense buffer gas system \cite{Vogl}. The spectral data shown in this work, as well as in the earlier redistribution laser cooling experiments \cite{Vogl}, use C-ring metal seals for the sealing of the sapphire optical cell windows of the high temperature pressure cells. A technical difficulty here were leakage problems during the heating up of the cells, which was most problematic for the collisional laser cooling experiments, which require a very high purity of the gases to suppress inelastic quenching effects. We have recently found suppliers that are capable of delivering flanges with welded sapphire windows, which are expected to benefit future experiments on this system. We point out that a redistribution laser cooling of molecules that are gaseous at room temperature could in future allow for the optical refrigeration of macroscopic gas samples directly starting from room temperature. The absorption wavelengths of electronic transitions of such molecules typically are in the ultraviolet spectral regime, with attractive systems e.g. being mixtures of the acetylene or the formaldehyde molecule with a noble buffer gas. The electronic transition wavelengths of those molecular systems are above 200 nm wavelength, i.e. are reachable with since recently commercially available high power diode laser based frequency-converted cw laser sources. Prospects of the redistribution laser cooling technique include the supercooling of gases below the homogeneous nucleation temperature \cite{Debenedetti} and optical chillers \cite{Bahae}.

\section*{Acknowledgments}

This work was supported by the Deutsche Forschungsgemeinschaft, grant No: We 1748-15.

\section*{References}	

\bigskip

\bibliography{master}   
\bibliographystyle{spiebib}   



\end{document}